# ON THE FLOW CURVE OF COLLOIDS PRESENTING SHEAR-INDUCED PHASE TRANSITIONS

**Daniel Quemada,**[1,*] **Claudio L. A. Berli**[2]

[1] *Groupe de Rhéologie, Matière et Systèmes Complexes,*
*Univ. Paris-Diderot (P7), Bât. Condorcet, MSC, case 7056 − 75205 Paris Cedex-13, France.*
[2] *INTEC (UNL-CONICET), Güemes 3450, 3000, Santa Fe, Argentina, and*
*Departamento de Física, FBCB, UNL, El Pozo, 3000, Santa Fe, Argentina.*

**Abstract:** This work deals with the evaluation of the flow curve of colloidal systems that develop fluid phases with different mechanical properties, namely shear-banding fluids. The problem involved is that, as different fluid phases coexist in the flow domain of the rheometric cell, measured data cannot be directly converted into rheometric functions. In order to handle this problem, a shear stress vs. shear rate constitutive relation is introduced to interpret the steady state flow curves. The relation derives from a phenomenological description of structural changes, and involves the possibility of multivalued shear rates under a given shear stress. Numerical predictions satisfactorily match up to experimental data of wormlike micellar solutions. A crucial aspect is the adequate computation of the shear rate function from raw data measured in the rheometric cell.

## 1. Introduction

During the last decade, rheometric measurements in combination with optical techniques provided detailed information on the flow characteristics of systems that develop fluid phases with different mechanical properties, namely shear bands (Butler, 1999; Lerouge *et al.*, 2000; Berret, 2005; Callahan, 2006). In these studies, the shear rate $\dot{\gamma}$ can be regarded as a field variable, since it is used to induce either the formation or disruption of microstructures, thus transforming fluid phases. In particular, the cell of concentric cylinders (Couette flow) is widely used for these measurements, where rheometric data are angular velocity $\Omega$ and torque $M$. Typical results found in the analysis of wormlike micellar solutions (WMS) are represented in Figure 1 (Cappelaere *et al.*, 1997; Salmon *et al.*, 2003). Inset boxes schematically show the microstructural changes driven by shear flow, as determined by optical techniques. The transition from isotropic towards non-isotropic structures leads to the observation of *bands* in the flow domain of the cell. In steady state flow, the bands coexist in a certain range of $\Omega$, where the curve $M(\Omega)$ notably decreases the slope (Figure 1).

---

[*] Corresponding autor (e-mail : quemada@ccr. jussieu.fr )

In this context, the present work discusses some rheological aspects concerning the treatment of experimental data of shear banded flows in Couette cells. It is worth to observe that, in this rheometry, the functions of interest are related to measured quantities through implicit equations. More precisely, the shear rate function $\dot{\gamma}$ appears in the kernel of an integral equation, which yields an inverse problem. In situations where two o more phases coexist in the flow domain of the cell, sharp variations of the fluid velocity occur, and hence appropriate numerical procedures are required to process data $M$ vs. $\Omega$, taking into account that an accurate determination of $\dot{\gamma}$ is desired. In fact, the knowledge of the true shear rate attained in the cell is crucial to investigate the relation between mechanical and structural phenomena in complex fluids. In order to handle this problem, here we suggest the use of a shear stress vs. shear rate constitutive relation for inelastic fluids, which involves the possibility of *multivalued* shear rates under defined conditions of shear stress. The relation derives from a phenomenological description of structural changes driven by shear flow (Quemada, 1982; 1998), and properly describes the flow curves of fluids presenting shear-induced phase transitions.

The paper is organized as follows: In Section 2, the problem related to the determination of the shear rate function in Couette rheometry is briefly reviewed and the model proposed for shear-banded flows is described. Then both the calculation suggested (Section 3) and its application to experimental data of WMS (Section 4) are discussed.

**2. Theoretical concepts**

*2.1 Equations of Couette rheometry for monophasic flows.*

Determining the flow curve $\sigma(\dot{\gamma})$ of non-Newtonian fluids requires the knowledge of both the shear stress $\sigma$ and the shear rate $\dot{\gamma}$ at one place in the flow domain of the cell. In Couette rheometry, $\sigma$ and $\dot{\gamma}$ are functions of the radial coordinate $r$, and they are related to measured data $\Omega$ vs. $M$ through the following equations (Walters, 1975; Macosko, 1994),

$$\sigma(r) = \frac{M}{2\pi r^2 L}, \tag{1}$$

$$\Omega = \int_{\kappa R}^{R} \frac{\dot{\gamma}(r)}{r} dr, \tag{2}$$

where $R$ and $\kappa R$ are the radii of outer and inner cylinder, respectively, and $L$ is the cylinders height. It is readily seen that $\sigma(r)$ is obtained straightforwardly from Eq. (1). In contrast, extracting $\dot{\gamma}(r)$ requires inverting the integral of Eq. (2), the solution of which is not unique due to the scattering present in experimental data. This inverse problem is also designated ill-posed in the literature (Friedrich *et al.*, 1996; Berli and Deiber, 2001). Indeed, direct estimations of $\dot{\gamma}$ can be achieved only when the gap between inner and outer cylinders is very small, say $1 > \kappa > 0.99$. Under these circumstances, $\dot{\gamma}(r)$ is considered nearly uniform throughout the flow domain, and thus Eq. (2) gives $\dot{\gamma}_{ng} \approx \Omega \kappa/(1-\kappa)$, which is known as the *narrow-gap* solution for Couette viscometry (Walters, 1975; Macosko, 1994).

For wider gap widths, an approach commonly used in practice consists in introducing some prior information on the fluid, namely a constitutive relationship $\sigma(\dot{\gamma})$. For example, the Power Law (PL) model, $\sigma = m\dot{\gamma}^n$ ($n$ is the flow index and $m$ is consistency parameter) allows one to solve Eqs. (1) and (2) analytically to obtain the angular velocity of the inner cylinder,

$$\Omega = (1 - \kappa^{2/n}) \frac{n}{2} \left( \frac{\sigma_{\kappa R}}{m} \right)^{1/n} \tag{3}$$

where $\sigma_{\kappa R} = M/(2\pi \kappa^2 R^2 L)$ is the shear stress applied at this cylinder. Therefore, given a *single-phase* fluid that obeys PL model, the parameters $n$ and $m$ can be determined by fitting Eq. (3) to experimental data $\Omega$ vs. $\sigma_{\kappa R}$, and then the shear rate can be calculated at any place in the flow domain, for instance at the inner wall as $\dot{\gamma}_{\kappa R} = (\sigma_{\kappa R}/m)^{1/n}$. Further, one may also integrate $\dot{\gamma}(r)$ to obtain the fluid velocity $u(r)$ in the cell.

*2.2. Equations of Couette rheometry for shear-banding flows.*

Shear-banding flows involve an additional complexity: abrupt changes in the function $\dot{\gamma}(r)$ arise when two or more phases coexist in the flow domain of the cell.

Different theoretical models aimed to predict the occurrence of shear-banding have been discussed in the literature (Spenley *et al.*, 1993; Porte *et al.*, 1997; Dhont, 1999; Radulescu and Olmsted, 2000; Fielding, 2005). Although a complete description of the phenomenon is not available yet, there is consensus in the literature that the underlying constitutive curve of shear-banding fluids has the form shown in Figure 2. In Couette cells, the shear stress is maximum at the inner cylinder ($\sigma_{\kappa R}$), and decreases smoothly as $\sigma(r) = \sigma_{\kappa R}(\kappa R/r)^2$, to give the minimum value at the outer cylinder ($\sigma_R = \kappa^2 \sigma_{\kappa R}$). The simple scenario considered here is that, when the stress $\sigma_{\kappa R}$ reaches the value $\sigma^*$, a new phase with lower flow resistance develops from the inner cylinder, and consequently $\dot{\gamma}_{\kappa R}$ jumps from $\dot{\gamma}_1$ to $\dot{\gamma}_2$ (Figure 2). Further increase in $\sigma_{\kappa R}$ results in the growth of this new phase, with a thickness enhancement of the associated band, up to complete development of the same phase in the whole gap, as $\sigma_{\kappa R}$ reaches $\sigma^*/\kappa^2$.

For the purpose of describing the rheometric problem, the following assumptions are needed: *i*) shear-bands are stable and can coexist in steady state conditions; *ii*) the bands present a localized interface at a certain $r = r^*$, where the shear stress is $\sigma = \sigma^*$; *iii*) both shear stress $\sigma(r)$ and fluid velocity $u(r)$ are continuous at the interface. In principle, these assumptions agree with hypothesis and experimental data reported by several authors (Cates *et al.*, 1993; Cappelaere *et al.*, 1997; Radulescu and Olmsted, 2000; Salmon *et al.*, 2003). However, in relation with the interface (*ii*), very recent theoretical (Fielding, 2005) and experimental (Lerouge *et al.*, 2006) works suggest the existence of a region of instability between the bands in the vorticity direction. As a first approximation, here we assume a sufficiently narrow and flat interface between the bands.

Under these conditions, and considering a functionality $\sigma(\dot{\gamma})$ as that plotted in Figure 2, the rheometric problem can be formulated as follows (see also Radulescu and Olmsted, 2000),

$$\sigma_{\kappa R} \leq \sigma^*, \qquad \Omega = -\frac{1}{2}\int_{\sigma_{\kappa R}}^{\sigma_R} \frac{\dot{\gamma}^<(\sigma)}{\sigma} d\sigma; \qquad (4a)$$

$$\sigma_{\kappa R} \geq \sigma^* \geq \sigma_R, \qquad \Omega = -\frac{1}{2}\left\{\int_{\sigma_{\kappa R}}^{\sigma^*} \frac{\dot{\gamma}^<(\sigma)}{\sigma} d\sigma + \int_{\sigma^*}^{\sigma_R} \frac{\dot{\gamma}^>(\sigma)}{\sigma} d\sigma\right\}; \qquad (4b)$$

$$\sigma_R \geq \sigma^*, \qquad \Omega = -\frac{1}{2}\int_{\sigma_{\kappa R}}^{\sigma_R} \frac{\dot{\gamma}^>(\sigma)}{\sigma} d\sigma. \qquad (4c)$$

In these equations, $\dot{\gamma}^<$ and $\dot{\gamma}^>$ indicate shear rate values $\dot{\gamma} < \dot{\gamma}_1$ and $\dot{\gamma} > \dot{\gamma}_2$, respectively. These values must be obtained from a suitable constitutive equation $\sigma(\dot{\gamma})$, as it is explained below.

*2.3. Fluid model proposed for shear-banding flows.*

Firstly it should be mentioned that analytical solutions to Eqs. (4a)-(4c) are simply attained by introducing $\dot{\gamma}(\sigma)$ according the PL model. In this sense, each band must be regarded as a different PL fluid. This also implies that Eq. (3) could represent the low and high shear rate zones, with different parameters *m* and *n* for each zone. Instead, in order to interpret the whole flow curve with a unique set of rheological parameters, here we use the following constitutive relationship for inelastic fluids,

$$\sigma(\dot{\gamma}) = \eta_\infty \left( \frac{1 + t_c \dot{\gamma}}{(\eta_\infty/\eta_0)^{1/2} + t_c \dot{\gamma}} \right)^2 \dot{\gamma}. \qquad (5)$$

This phenomenological equation derives from a kinetic description of structural changes induced by shear (Quemada, 1982; 1998). It is assumed that, when a given shear stress is applied, a sort of order-disorder equilibrium is established, in which the forward (Brownian motion) and backward (shear-induced ordering) processes balance. Thus in Eq. (5), $\eta_0$ and $\eta_\infty$ are the limiting viscosities for $\dot{\gamma} \to 0$ and $\dot{\gamma} \to \infty$, respectively, and $t_c$ is a characteristic relaxation time. In particular, given an appropriate set of parameters, Eq. (5) predicts multiple values of $\dot{\gamma}$ for a given shear stress, as shown in Figure 2. It is interesting to observe however that the viscosity $\eta(\dot{\gamma}) = \sigma(\dot{\gamma})/\dot{\gamma}$ related to Eq. (5) is a purely monotonic function.

As seen in Figure 2, there is a range of $\sigma$ for which, in principle, $\dot{\gamma}$ could jump from the low-shear branch to the high-shear one. Nevertheless, a unique and reproducible value $\sigma^*$ is observed in experiments (Cappelaere *et al.*, 1997; Lu *et al.*, 2000, Salmon *et al.*, 2003). The selection of the shear stress $\sigma^*$ at which the new band develops is a crucial aspect in modeling shear-banding flows. Indeed, the determination of a criterion selection is still an open problem, and several authors discuss the mechanism to be applied in different constitutive models (for instance, Lu *et al.*, 2000, and references therein). In the present work, an *ad hoc* value $\sigma^*$ will be introduced in calculations to satisfy the experimental data.

## 3. Calculation procedure

The values $\dot{\gamma}^<(\sigma)$ and $\dot{\gamma}^>(\sigma)$ entering Eqs. (4a)-(4c) are obtained from Eq. (5), as numerical roots $\dot{\gamma}(\sigma)$ for a given set of known parameters ($\eta_0$, $\eta_\infty$, $t_c$). This task is carried out through a Newton-Raphson subroutine (Carnahan *et al.*, 1969), for around $10^4$ discrete values of $\sigma$ in the range $\sigma_R \leq \sigma \leq \sigma_{\kappa R}$. Then the integrals in Eqs. (4a)-(4c) are solved numerically by using the trapezoidal rule (Carnahan *et al.*, 1969), also with $10^4$ discrete intervals. This is carried out for an initial, tentative value $\sigma^*$, normally inferred from the experimental curve, which is then adjusted to provide the best representation of data.

One should underline that the present calculation takes into account the variations of $\dot{\gamma}^<(\sigma)$ and $\dot{\gamma}^>(\sigma)$ in each band. Hence, it differs noticely from current asumptions made in number of studies (for example, Georgiou and Vlassopoulos, 1998; Salmon *et al.*, 2003; Drappier, 2004). In these studies, $\dot{\gamma}^<$ and $\dot{\gamma}^>$ are taken as the limits $\dot{\gamma}_1$ and $\dot{\gamma}_2$ of the plateau $\sigma(\dot{\gamma}) = \sigma^*$, hence are constants entering the "lever rule" $e\dot{\gamma} = e_1\dot{\gamma}_1 + e_2\dot{\gamma}_2$, where $e_1$ and $e_2$ are the band thicknesses, $e = R - \kappa R$ is the gap width, and $\dot{\gamma}$ is the "measured" shear rate. Such constant values are not observed in measured velocity profiles (see, for example, Figure 2 in Salmon *et al.*, 2003). Moreover, one may expect that the larger the gap thickness *e*, the higher the differences with these constant values.

Finally, if the parameters ($\eta_0$, $\eta_\infty$, $t_c$) of a given fluid are known, Eqs. (4a)-(5) predict the values $\Omega$ vs. $\sigma$ (or *M*) to be obtained in an experiment in which the requirements to accomplish a viscometric flow are satisfied (steady state, no-slip at the walls, end effects negligible, isothermal flow), as well as the assumptions (*i-iii*) made above. This calculation may be designated *direct calculation*, and it will be illustrated in detail below. Previously, it is relevant to mention that a more challenging problem is the *inverse calculation*: determining the values of the parameters ($\eta_0$, $\eta_\infty$, $t_c$) from the curve of raw data $\Omega$ vs. $\sigma$ (or *M*), and then using them to plot the flow curve $\sigma(\dot{\gamma})$ in the appropriate range of shear rates. The implementation of this task requires further efforts, as the minimization problem involved cannot be solved with standard mathematical software (see, for instance, Berli and Deiber, 2001).

## 4. Results and discussion

This section illustrates the applicability of the model to interpret rheometric data of shear-banding systems. In particular, experimental data of WMS published in the literature are considered. These data, which were reported as $\sigma_{\kappa R}$ vs. $\dot{\gamma}_{ng}$ for two surfactant systems, CTAB (Cappelaere *et al.*, 1997) and CPCl-NaSal (Salmon *et al.*, 2003), are presented here as $\sigma_{\kappa R}$ vs. $\Omega$ in Figure 3.

The first step consists in confronting Eq. (5) to data $\sigma$ vs. $\dot{\gamma}$, for which one firstly needs accurate values of $\dot{\gamma}$. For this purpose, we select experimental data of Figure 3 in the intervals where the fluid is *monophasic* only. Thus the low and high shear zones were analyzed independently one another, under the assumption that each zone obeys to PL model. The parameters $n$ and $m$ for each zone were obtained by fitting Eq. (3) to data $\sigma_{\kappa R}(\Omega)$, and they are reported in Table 1. Then the shear rate $\dot{\gamma}_{\kappa R}$ was calculated, such as it was described in the previous section. Results are presented in Figure 4 (symbols). Also in this figure, full lines represent the prediction of Eq. (5), with the values of $\eta_0$, $\eta_\infty$, and $t_c$ that fit this data, also reported in Table 1. It is observed that the model describes satisfactorily the flow curve in the full range of shear rates, by predicting an intermediate *multivalued* zone.

In order to cross-check these results, and having the parameters $\eta_0$, $\eta_\infty$, and $t_c$ that characterize the fluid, we finally carried out the *direct* calculation. That is, Eqs. (4a)-(5) were solved numerically, and the resulting function $\Omega(\sigma_{\kappa R})$ was matched to the respective experimental curve. This is actually done in Figure 3, where full lines are the numerical predictions with the values of $\sigma^*$ indicated in the figure caption. A remarkable agreement is observed in the full range of the experimental data. Furthermore, it is worthy of note that the values of $\sigma^*$ used here agree with those previously reported (Cappelaere *et al.*, 1997; Salmon *et al.*, 2003).

The fact that $\sigma^*$ values are rather close to the respective minimum $\sigma_{min}$ of each flow curve $\sigma(\dot{\gamma})$ in both set of data (Figure 4) could be tentatively interpreted as follows. As $\sigma$ increases from 0, one may expect that the flow in the gap should become unstable as soon as $\sigma_{\kappa R}$ reaches $\sigma_{min}$: at this moment, the number of roots of the equation $\dot{\gamma}(\sigma_{\kappa R})$ then passes from 1 to 3, the intermediate one corresponding to an unstable state. Moreover, it seems plausible to relate the slight difference between $\sigma^*$ and $\sigma_{min}$ (with $\sigma^* > \sigma_{min}$) to some delay required to built the bands after $\sigma_{\kappa R}$ has reached $\sigma_{min}$. Justifying such expectations will demand further research.

## 5. Concluding remarks and future directions

The present paper basically discusses the prediction of rheometric data of shear-banding flows by using a constitutive model for flow curve of the fluid. It may be remarked that the success of calculations suggested relies on both the introduction of a suitable model for the fluid, and the adequate computation of the shear rate function from raw data measured in the rheometric cell. Indeed, when fluids presenting flow indexes *n* as low as 0.2 are studied, an inappropriate estimation of the shear rate leads to considerable errors, notably when two phases coexist in the flow domain.

Another crucial aspect in shear banding flows is the modeling of unsteady rheometric data; more precisely, shear stress vs. time curves obtained after the sudden inception of a given shear rate. Typical responses observed are stress overshoot, dumped oscillations and simple relaxation to the steady state (Lerouge *et al.*, 2006). Theoretical descriptions of these experiments are rather demanding, as different time-dependent phenomena are involved, mainly the transient of the apparatus, relaxation of the fluid microstructrure, and viscoelasticity of the fluid. The structural model, from which Eq. (5) derives, is also able to interpret unsteady curves on the base of kinetic processes (Quemada, 2006). Structural model predictions are aimed to be compared with those from Johnson-Segalman model, which is the most employed constitutive relation to describe shear-banding fluids (Radulescu and Olmsted, 2000; Wilson and Fielding, 2006). These topics are to be considered in a future work.

**Acknowledgements**

C. L. A. B. acknowledges the financial aid received from CONICET and SEPCYT-FONCyT (PICT 09-14732), Argentina.

**Table 1**. Model parameters of constitutive fluid models concerning data in Figures 3 and 4.

| Model | Data/Zone | Parameter | System | |
|---|---|---|---|---|
| | | | CPCl/NaSal | CTAB |
| Eq. (3) | $\sigma(\Omega)$/Low | $m$ (Pa s$^n$) | 35 | 6.7 |
| | | $n$ | 0.7 | 0.84 |
| | $\sigma(\Omega)$/High | $m$ (Pa s$^n$) | 34.3 | 4.96 |
| | | $n$ | 0.19 | 0.31 |
| Eq. (5) | $\sigma(\dot{\gamma})$/Full | $\eta_0$ (Pa s) | 63.6 | 7.8 |
| | | $\eta_\infty$ (Pa s) | 0.64 | 0.024 |
| | | $t_c$ (ms) | 32.3 | 2.77 |

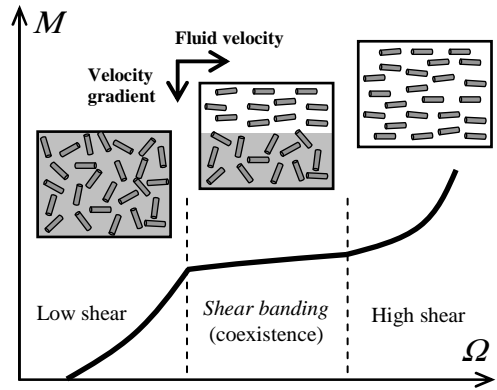

**Figure 1.** Curve of rheometric data (arbitrary drawing) typically found in WMS studied by Couette flow. The insets are highly schematic representations of the solution structure in different shear rates zones. At a given shear stress, a new fluid phase (*band*) develops as micelles align in the flow direction.

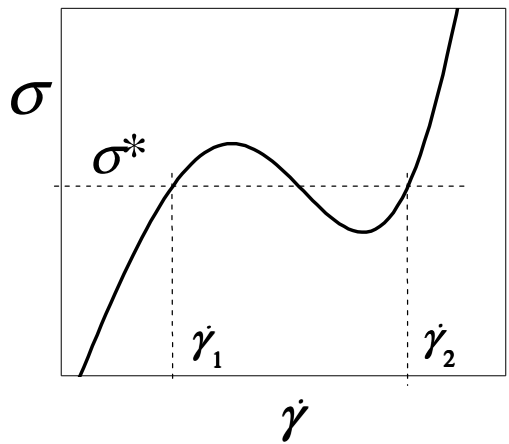

**Figure 2.** Shear stress as a function of shear rate for shear-thinning, -banding fluids (schematic draw, arbitrary units).

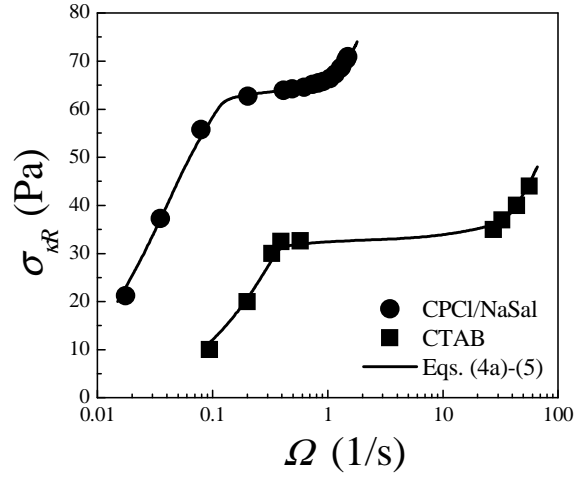

**Figure 3.** Shear stress as a function of angular velocity for different WMS. Symbols represent two examples of experimental data reported in the literature: (,) Salmon *et al.*, 2003; (!) Cappelaere *et al.*, 1997. Full lines are the predictions of Eqs. (4a)-(5), with the values of $\eta_0$, $\eta_\infty$, and $t_c$ reported in Table 1. In addition, $\sigma^* = 62$ Pa for CPCl/NaSal, and $\sigma^* = 32.5$ Pa for CTAB.

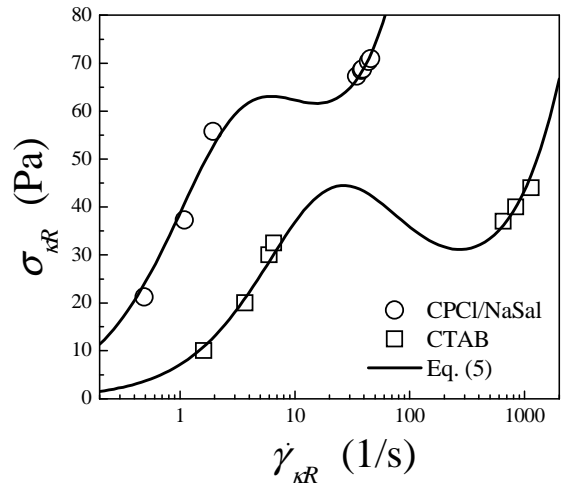

**Figure 4.** Shear stress as a function of shear rate, corresponding to the systems presented in Figure 3. Symbols are the values obtained from experimental data $\sigma_{\kappa R}(\Omega)$ in the shear zones where the fluid is monophasic (see text for details). Full lines are the predictions of Eq. (5), with the values of $\eta_0$, $\eta_\infty$, and $t_c$ reported in Table 1.